# An End-to-End Probabilistic Network Calculus with Moment Generating Functions

Markus Fidler
Department of Electrical and Computer Engineering
University of Toronto, Ontario, Canada
fidler@ieee.org

*Abstract*—Network calculus is a min-plus system theory for performance evaluation of queuing networks. Its elegance stems from intuitive convolution formulas for concatenation of deterministic servers. Recent research dispenses with the worst-case assumptions of network calculus to develop a probabilistic equivalent that benefits from statistical multiplexing. Significant achievements have been made, owing for example to the theory of effective bandwidths, however, the outstanding scalability set up by concatenation of deterministic servers has not been shown.

This paper establishes a concise, probabilistic network calculus with moment generating functions. The presented work features closed-form, end-to-end, probabilistic performance bounds that achieve the objective of scaling linearly in the number of servers in series. The consistent application of moment generating functions put forth in this paper utilizes independence beyond the scope of current statistical multiplexing of flows. A relevant additional gain is demonstrated for tandem servers with independent cross-traffic.

## I. INTRODUCTION

Network calculus [1], [2] is a theory of deterministic queuing systems that is based on the early $(\sigma, \rho)$-calculus for network delay [3] and on the work on generalized processor sharing in [4]. Founded on min-plus algebra [5] it relates to classical system theory, where output and concatenation of systems can be derived by intuitive convolution formulas [6], [7], [8], [9], [10]. Owing to these network calculus allows for an efficient analysis of networks of queues. It facilitates the derivation of worst-case backlogs and delays by applying deterministic upper envelopes on traffic arrivals and lower envelopes on the offered service, so-called arrival and service curves [6], [4].

Although the fundamental performance bounds provided by deterministic network calculus were proven to be tight, that is they are attained for certain sample paths [2], the occurrence of such worst-case events is usually rare, especially if aggregated traffic is considered. Statistical multiplexing of independent flows is known to smooth out burstiness with a high probability, whereas bursts are cumulated by deterministic network calculus, resulting in pessimistic bounds and overestimation of resource requirements. This has motivated considerable research in recent time aiming at a probabilistic equivalent which efficiently utilizes statistical multiplexing while preserving powerful concatenation properties.

This work was supported by the DFG under an Emmy Noether grant, by the Institute Mittag-Leffler under the program on Queueing and Teletraffic Theory, by the NSF under grant CNS-0435061, and by two NSERC Discovery grants.

Introducing independence of flows to the framework of network calculus tightens assumptions, such that traffic sources must not be adversarial jointly, although independent increments are not assumed and sources may still create worst-case traffic patterns on their own. Adversarial sources and the worst-case patterns generated by extremal shape controlled traffic were subject to detailed investigations [11]. The performance of multiplexing leaky bucket and peak rate constrained on-off sources was investigated in [12] and further elaborated in [13], where buffer and server were decoupled. General models for regulated traffic were investigated in [14]. A corresponding framework for bufferless multiplexing was derived in [15] and for buffered multiplexing in [16]. Improvements and a generalization of the bounds in [16], [17] were provided in [18].

A more generic class of traffic arrivals was investigated in [19] where an exponential bound on the burstiness was applied. The approach was extended to the concept of stochastically bounded burstiness in [20] and generalized in [21]. Related models were also used in [22], [23], [1], [24], where the pioneering work in [23], [1] carried the deterministic $(\sigma, \rho)$-calculus from [3] forward to a stochastic setting using $(\sigma(\theta), \rho(\theta))$-envelopes of the moment generating function (MGF) of arrival processes. Backlog, delay, and output bounds for work-conserving links were derived from Chernoff's theorem in [23], [1]. A related theory for intree networks was developed in [25] using stochastic ordering.

Generally, probabilistic backlog, delay, and output bounds for a single server facilitate the iterative application to networks of queues, using respective output bounds as input bounds for subsequent downstream servers and adding per-server delay and backlog bounds up. Unfortunately, this approach results in end-to-end performance bounds which are loose and decay rapidly with the number of servers that are traversed [26]. Few models are, however, known that allow concatenating probabilistic service curves to derive end-to-end probabilistic network models.

An important continuation of the work in [23] was put forward in [1] applying the more general $f(\theta, t)$-upper constraint for traffic sequences and a corresponding lower constraint for the service offered by servers with time varying capacity. The proposed framework provides calculation formulas for stochastic backlog and delay bounds as well as the fundamental concatenation property. Related results for max-plus systems [5] were derived in [27] and applied to tandem queues using recursive equations to describe departure times which were then solved in matrix formulation.

A significant, alternative step towards a probabilistic, end-to-end network calculus was presented in [28], where the concept of effective envelopes, which are violated at most with a defined probability, was derived using MGFs and the Chernoff bound. Related theories use the concept of statistical envelopes [29] and apply second order statistics [30]. The concept of effective envelopes was further developed in [31], [32] and [33] where the relation to the theory of effective bandwidths, for example [1], [34], was shown. MGFs are used extensively by the theory of effective bandwidths from where a variety of useful traffic models are known, see for example [1], [34], [28], [23], [16], [33] and references therein.

Effective envelopes are conceptually close to deterministic envelopes as applied by network calculus, resulting in an intuitive continuation including the important concatenation theorem. However, the derivation of violation probabilities for performance bounds as well as for concatenated effective service curves has proven to be hard [33] and requires additional assumptions, for example bounds on the duration of busy periods [33] or explicit dropping of traffic that violates a certain delay bound [35]. Once effective envelopes are fixed no further statistical multiplexing gain is accessible and worst-case assumptions apply, such that independence is not required any longer.

In recent work [26] the Landau notation $\mathcal{O}$ was introduced to the field of network calculus to analyze the scaling of end-to-end performance bounds in the number of servers in series $n$. This important aspect provides comparability of different models and improves their understanding. The standards set by deterministic network calculus are a scaling in $\mathcal{O}(n^2)$ for bounds that are derived iteratively compared to $\mathcal{O}(n)$ obtained from end-to-end convolved service curves [2]. In [26] an outstanding scaling of statistical bounds in $\mathcal{O}(n \log n)$ was achieved without assuming statistical independence, where rate correction terms were introduced which dispense with busy period bounds.

Until now it has, however, not been proven that end-to-end probabilistic performance bounds can actually scale as well as deterministic bounds. Based on the pioneering stochastic $(\sigma(\theta), \rho(\theta))$-calculus [23] and following the suggestion in [1] we develop an intuitive, system-theoretic formulation of a probabilistic network calculus with MGFs which achieves the target scaling in $\mathcal{O}(n)$ without requiring assumptions beyond existence of respective MGFs and statistical independence.

The paper is organized as following: We derive corresponding MGFs of relevant min-plus operations in Sect. II. In Sect. III we develop a probabilistic network calculus based on the results from Sect. II. The scaling properties of performance bounds are proven in Sect. IV and numerical results are shown in Sect. V. Throughout the paper we use a discrete time model $t \in \mathbb{N}_0 = \{0, 1, 2, \dots\}$.

## II. MOMENT GENERATING FUNCTIONS

The particular strength of network calculus is the efficient analysis of servers in series. Tandem servers can be concatenated using min-plus convolution and performance bounds can be derived based on min-plus de-convolution. Min-plus convolution of real-valued, bivariate functions $x(s,t)$ and $y(s,t)$ [1] and a corresponding min-plus de-convolution are defined for $t \geq s \geq 0$ as

$$(x \otimes y)(s,t) = \inf_{\tau \in [s,t]}[x(s,\tau) + y(\tau,t)]$$
$$(x \oslash y)(s,t) = \sup_{\tau \in [0,s]}[x(\tau,t) - y(\tau,s)]. \quad (1)$$

Note that the defined operators $\otimes$ and $\oslash$ are not commutative.

The applicability of network calculus suffers from the worst-case modelling that is used. The deterministic view totally neglects statistical multiplexing which, however, potentiates significant resource savings. Related aspects are well understood for single servers and the theory of effective bandwidths provides a general traffic model which makes the statistical multiplexing gain among independent, multiplexed flows accessible. Effective bandwidths resort to MGFs which for a random variable $X$ are defined for any $\theta$ as

$$\mathsf{M}_X(\theta) = \mathsf{E}e^{\theta X}$$

where $\mathsf{E}$ is the expectation of its argument. In the sequel we use the notation $\overline{\mathsf{M}}_X(\theta) = \mathsf{M}_X(-\theta) = \mathsf{E}e^{-\theta X}$.

MGFs exhibit a number of useful properties: Given two constants $c_1$ and $c_2$ it is known that $\mathsf{M}_{c_1 + c_2 X}(\theta) = e^{\theta c_1}\mathsf{M}_X(c_2 \theta)$. Further on, for two random variables $X$ and $Y$ and for $\theta \geq 0$ it can be easily verified that

$$\mathsf{M}_{\min[X,Y]}(\theta) \leq \min[\mathsf{M}_X(\theta), \mathsf{M}_Y(\theta)]$$
$$\overline{\mathsf{M}}_{\max[X,Y]}(\theta) \leq \min[\overline{\mathsf{M}}_X(\theta), \overline{\mathsf{M}}_Y(\theta)] \quad (2)$$

since for $\theta \geq 0$ it holds that $\mathsf{M}_{\min[X,Y]}(\theta) \leq \mathsf{M}_X(\theta)$ as well as $\mathsf{M}_{\min[X,Y]}(\theta) \leq \mathsf{M}_Y(\theta)$. For the sum of independent random variables $X$ and $Y$ it is well-known for any $\theta$ that

$$\mathsf{M}_{X+Y}(\theta) = \mathsf{M}_X(\theta)\mathsf{M}_Y(\theta)$$
$$\mathsf{M}_{X-Y}(\theta) = \mathsf{M}_X(\theta)\overline{\mathsf{M}}_Y(\theta). \quad (3)$$

Lastly Chernoff's theorem builds on MGFs to estimate the violation probability of an effective bound $x$ for a random variable $X$. For any $x$ and all $\theta \geq 0$ it is known that

$$\mathsf{P}\{X \geq x\} \leq e^{-\theta x}\mathsf{E}e^{\theta X}.$$

Throughout the remainder of this section we proof an important connection between min-plus algebra and conventional algebra established by MGFs [27], [1]. We will show that MGFs transform min-plus convolution and de-convolution into corresponding operators $*$ and $\circ$ in conventional algebra which for real-valued, bivariate functions $x(s,t)$ and $y(s,t)$ are defined for $t \geq s \geq 0$ as

$$(x * y)(s,t) = \sum_{\tau=s}^{t} x(s,\tau)y(\tau,t)$$
$$(x \circ y)(s,t) = \sum_{\tau=0}^{s} x(\tau,t)y(\tau,s). \quad (4)$$

We will sometimes use the operators $\oslash$ and $\circ$ for $s \geq t \geq 0$ whereupon the supremum respectively the sum is evaluated for all $\tau \in [0,t]$ which for notational convenience is omitted in (1) and (4).

The following lemmas act on a suggestion made in [1] and are in accordance with a similar investigation of max-plus systems in [27].

**Lemma 1 (Moment Generating Functions of $\otimes$ and $\oslash$)**
Let $X(s,t)$ and $Y(s,t)$ be independent random processes. The MGF of the min-plus convolution respectively the MGF of the min-plus de-convolution of $X(s,t)$ and $Y(s,t)$ are upper bounded according to

$$\overline{\mathsf{M}}_{(X \otimes Y)}(\theta,s,t) \leq (\overline{\mathsf{M}}_X(\theta) * \overline{\mathsf{M}}_Y(\theta))(s,t)$$
$$\mathsf{M}_{(X \oslash Y)}(\theta,s,t) \leq (\mathsf{M}_X(\theta) \circ \overline{\mathsf{M}}_Y(\theta))(s,t).$$

*Proof:* The MGF of the min-plus convolution of $X(s,t)$ and $Y(s,t)$ is

$$\overline{\mathsf{M}}_{(X \otimes Y)}(\theta,s,t) = \mathsf{E} e^{-\theta \inf_{\tau \in [s,t]}[X(s,\tau)+Y(\tau,t)]}.$$

An upper bound on the MGF follows for any $\theta$ as

$$\overline{\mathsf{M}}_{(X \otimes Y)}(\theta,s,t) \leq \mathsf{E} \sup_{\tau \in [s,t]} \left[ e^{-\theta(X(s,\tau)+Y(\tau,t))} \right]$$
$$\leq \mathsf{E} \sum_{\tau=s}^{t} e^{-\theta(X(s,\tau)+Y(\tau,t))}$$
$$= \sum_{\tau=s}^{t} \mathsf{E} e^{-\theta(X(s,\tau)+Y(\tau,t))}.$$

Applying (3) and (4) completes the proof. The proof for min-plus de-convolution $\oslash$ is an immediate variation. ∎

Unlike [28], [33], where MGFs and the Chernoff bound are used to derive effective envelopes before min-plus convolution and de-convolution are utilized, we use only MGFs and apply the Chernoff bound in a final step to derive bounds on the results of min-plus convolution and de-convolution. This approach maintains random processes as long as possible, which in case of statistical independence is beneficial for the quality of performance bounds.

**Lemma 2 (Envelopes for $\otimes$ and $\oslash$)** Let $X(s,t)$ and $Y(s,t)$ be independent random processes. Effective lower envelopes on the min-plus convolution respectively upper envelopes on the min-plus de-convolution of $X(s,t)$ and $Y(s,t)$ that hold at least with probability $1 - \varepsilon$ where $\varepsilon \in (0,1]$ are given by

$$\mathsf{P}\Big\{(X \otimes Y)(s,t) \geq$$
$$\sup_{\theta \in (0,\infty)} \Big[\frac{1}{\theta}\big(\ln \varepsilon - \ln(\overline{\mathsf{M}}_X(\theta) * \overline{\mathsf{M}}_Y(\theta))(s,t)\big)\Big]\Big\} \geq 1-\varepsilon,$$

$$\mathsf{P}\Big\{(X \oslash Y)(s,t) \leq$$
$$\inf_{\theta \in (0,\infty)} \Big[\frac{1}{\theta}\big(\ln(\mathsf{M}_X(\theta) \circ \overline{\mathsf{M}}_Y(\theta))(s,t) - \ln \varepsilon\big)\Big]\Big\} \geq 1-\varepsilon.$$

*Proof:* Boole's inequality yields

$$\mathsf{P}\{(X \otimes Y)(s,t) \leq z(s,t)\}$$
$$= \mathsf{P}\Big\{\inf_{\tau \in [s,t]}[X(s,\tau)+Y(\tau,t)] \leq z(s,t)\Big\}$$
$$= \mathsf{P}\Big\{\bigcup_{\tau=s}^{t}\{X(s,\tau)+Y(\tau,t) \leq z(s,t)\}\Big\}$$
$$\leq \sum_{\tau=s}^{t} \mathsf{P}\{X(s,\tau)+Y(\tau,t) \leq z(s,t)\}.$$

Applying Chernoff's bound it follows for all $\theta \geq 0$ that

$$\mathsf{P}\{(X \otimes Y)(s,t) \leq z(s,t)\}$$
$$\leq \sum_{\tau=s}^{t} \mathsf{P}\{-X(s,\tau)-Y(\tau,t) \geq -z(s,t)\}$$
$$\leq \sum_{\tau=s}^{t} e^{\theta z(s,t)} \mathsf{E} e^{-\theta(X(s,\tau)+Y(\tau,t))}.$$

Setting the right hand side equal to $\varepsilon$ we can solve for $z(s,t)$ and obtain

$$z(s,t) = \frac{1}{\theta}\Big(\ln \varepsilon - \ln\Big(\sum_{\tau=s}^{t} \mathsf{E} e^{-\theta(X(s,\tau)+Y(\tau,t))}\Big)\Big)$$

for all $\theta \in (0,\infty)$ where we choose the optimal $\theta$. The proof for min-plus de-convolution $\oslash$ is an immediate variation. ∎

For completeness we note that related formulas as provided by Lemma 1 and Lemma 2 can be derived without assuming independence in which case these operations, however, become costly. From Hölder's inequality $\mathsf{E}|AB| \leq (\mathsf{E}|A|^\kappa)^{1/\kappa}(\mathsf{E}|B|^\nu)^{1/\nu}$ where $\kappa, \nu > 1$ and $1/\kappa + 1/\nu = 1$ it follows for the sum of two random variables $X$ and $Y$ with $A = e^{\theta X}$ and $B = e^{\pm \theta Y}$ for all $\theta$ that

$$\mathsf{M}_{X+Y}(\theta) \leq (\mathsf{M}_X(\kappa\theta))^{1/\kappa}(\mathsf{M}_Y(\nu\theta))^{1/\nu}$$
$$\mathsf{M}_{X-Y}(\theta) \leq (\mathsf{M}_X(\kappa\theta))^{1/\kappa}(\overline{\mathsf{M}}_Y(\nu\theta))^{1/\nu}$$

which replaces (3) whenever independence cannot be assumed. Related theories which do not require independence are shown in [19], [20] and are significantly further advanced for effective envelopes [28], [31], [26], [33], [32], whereas this work focuses on exploiting statistical independence.

The dual operators $*$ and $\circ$ form the basis for an intuitive probabilistic network calculus with MGFs which features the efficient utilization of statistical multiplexing within a framework for effective concatenation of tandem servers. Note that Lemma 1 can be applied iteratively to more than one min-plus convolution or de-convolution, for example an upper bound on the MGF of $(X \oslash (Y_1 \otimes \cdots \otimes Y_n))(s,t)$ is given by $(\mathsf{M}_X(\theta) \circ (\overline{\mathsf{M}}_{Y_1}(\theta) * \cdots * \overline{\mathsf{M}}_{Y_n}(\theta)))(s,t)$. This property is an essential prerequisite for performance analysis of tandem servers as shown in the following sections.

## III. PROBABILISTIC NETWORK CALCULUS

In this section we recall fundamental properties of network calculus, see for example [1], and apply the results from Sect. II to derive probabilistic performance bounds.

Arrival and departure processes are described by real-valued, cumulative functions $A(0,t)$ and $D(0,t)$ respectively which represent the amount of data seen in the interval $(0,t]$. We assume that there are no arrivals in the interval $(-\infty, 0]$. Clearly $A(0,t)$ and $D(0,t)$ are nonnegative and increasing in $t$. The amount of data seen in an interval $(s,t]$ is denoted by $A(s,t) = A(0,t) - A(0,s)$ and $D(s,t) = D(0,t) - D(0,s)$, where $A(s,t)$ and $D(s,t)$ are nonnegative, increasing in $t$, decreasing in $s$, and $A(t,t) = 0$ and $D(t,t) = 0$ for all $t$.

**Definition 1 (Dynamic Server)** Assume $A(0,t)$ and $D(0,t)$ are the arrival respectively departure process of a lossless server. Let $S(s,t)$ for $t \geq s \geq 0$ be a random process that is nonnegative and increasing in $t$. The server is called a dynamic server $S(s,t)$ if for any fixed sample path it holds for all $t \geq 0$ that

$$D(0,t) \geq (A \otimes S)(0,t).$$

The definition of dynamic server is according to [1], [36], [37]. It extends the framework of deterministic network calculus using bivariate functions. Note that while $A(s,t)$ is defined as $A(s,t) = A(0,t) - A(0,s)$ similar assumptions are not generally made for $S(s,t)$.

**Example 1 (Work-Conserving Server)** Consider a work-conserving server with arrival and departure process $A(0,t)$ and $D(0,t)$ respectively. For any $t \geq 0$ fix $\tau = \sup[s \in [0,t] : D(0,s) = A(0,s)]$, that is $\tau$ is the beginning of the last backlogged period before $t$ if any or otherwise $\tau = t$. Let $S(\tau, t)$ be a random process that denotes the service offered by the server in the interval $(\tau, t]$ under the condition that $\tau$ is the beginning of the last backlogged period before $t$. Assume that $S(\tau, t)$ is nonnegative, increasing in $t$ and $S(\tau, \tau) = 0$. From the work-conserving property we have for any fixed sample path that $D(0,t) = D(0,\tau) + S(\tau, t)$, that is the server is non-idling and uses the entire available service $S(\tau, t)$ to serve backlogged data. With $D(0,\tau) = A(0,\tau)$ it follows that $D(0,t) = A(0,\tau) + S(\tau, t)$ whereby Def. 1 is fulfilled.

Performance bounds for dynamic servers have been obtained in [1], [36], [37] where it has also been shown that dynamic servers in series can be effectively transformed into an equivalent, single dynamic server. Thus, known performance bounds extend immediately to tandem servers. For completeness Th. 1 and Th. 2 rephrase these results formally and show corresponding MGFs respectively probabilistic bounds. In the sequel upper indices indicate servers respectively belonging arrival and departure processes of a server.

**Theorem 1 (Concatenation)** *Consider two dynamic servers $S^1(s,t)$ and $S^2(s,t)$ in series. There exists an equivalent, single dynamic server $S(s,t)$ for $t \geq s \geq 0$ where*

$$S(s,t) = (S^1 \otimes S^2)(s,t).$$

*Assume the service offered by each of the servers has MGF $\overline{\mathsf{M}}_{S^1}(\theta, s, t)$ respectively $\overline{\mathsf{M}}_{S^2}(\theta, s, t)$. Under the assumption of statistical independence of $S^1(s,t)$ and $S^2(s,t)$ the MGF of the equivalent, single dynamic server is upper bounded for $t \geq s \geq 0$ according to*

$$\overline{\mathsf{M}}_S(\theta, s, t) \leq (\overline{\mathsf{M}}_{S^1}(\theta) * \overline{\mathsf{M}}_{S^2}(\theta))(s,t).$$

*Proof:* Let $A^i(0,t)$ and $D^i(0,t)$ be the arrival respectively departure process of server $i$, where $A^i(0,t) = D^{i-1}(0,t)$. For any fixed sample path it follows from Def. 1 for all $t \geq 0$ that

$$\exists \tau \in [0,t] : D^2(0,t) - A^2(0,\tau) \geq S^2(\tau, t).$$

In the same way it holds for any $\tau \geq 0$ that

$$\exists s \in [0,\tau] : D^1(0,\tau) - A^1(0,s) \geq S^1(s,\tau).$$

Taking the sum yields for all $t \geq 0$ that

$$\exists \tau \in [0,t], \exists s \in [0,\tau] : D^2(0,t) - A^1(0,s) \geq S^1(s,\tau) + S^2(\tau,t)$$

where the right hand side is the min-plus convolution of $S^1$ and $S^2$ which proves the first part. An upper bound on the MGF of min-plus convolution of two statistically independent random processes $(S^1 \otimes S^2)(s,t)$ follows from Lemma 1 as $(\overline{\mathsf{M}}_{S^1}(\theta) * \overline{\mathsf{M}}_{S^2}(\theta))(s,t)$. ∎

Two performance measures, namely backlog and delay as defined below, are of particular interest in the context of networking and networked applications.

**Definition 2 (Backlog and Delay)** Let $A(0,t)$ and $D(0,t)$ be the arrival respectively departure process of a lossless server. The backlog at time $t \geq 0$ is

$$b(t) = A(0,t) - D(0,t).$$

Assuming first-in first-out ordering the delay at time $t \geq 0$ is defined as

$$d(t) = \inf[s \geq 0 : A(0,t) - D(0,t+s) \leq 0].$$

**Theorem 2 (Backlog, Delay, and Output Bounds)**
*Consider a dynamic server $S(s,t)$ with arrival process $A(s,t)$. The backlog at time $t \geq 0$ is upper bounded according to*

$$b(t) \leq (A \oslash S)(t,t).$$

*Assuming first-in first-out scheduling the delay at time $t \geq 0$ is upper bounded according to*

$$d(t) \leq \inf[s \geq 0 : (A \oslash S)(t+s,t) \leq 0].$$

*Note that the delay bound formulation using min-plus deconvolution extends the domain of the definition in (1) which is accounted for by taking the supremum over the adapted interval $[0,t]$, that is $(A \oslash S)(t+s,t) = \sup_{\tau \in [0,t]}[A(\tau,t) - S(\tau, t+s)]$.*

*The departure process $D(s,t)$ is upper bounded for any $t \geq s \geq 0$ according to*

$$D(s,t) \leq (A \oslash S)(s,t).$$

Assume the arrival process has MGF $\mathsf{M}_A(\theta, s, t)$ and the service offered by the server has MGF $\overline{\mathsf{M}}_S(\theta, s, t)$. Under the assumption of statistical independence of $A(s, t)$ and $S(s, t)$ an upper bound on $(A \oslash S)(s, t)$ with violation probability $\varepsilon \in (0, 1]$ is given by

$$\mathsf{P}\Big\{(A \oslash S)(s, t) \leq$$
$$\inf_{\theta \in (0, \infty)} \left[\frac{1}{\theta}(\ln(\mathsf{M}_A(\theta) \circ \overline{\mathsf{M}}_S(\theta))(s, t) - \ln \varepsilon)\right]\Big\} \geq 1 - \varepsilon$$

which provides an upper bound on $(A \oslash S)(s, t)$ that is violated at most with probability $\varepsilon$ and in turn establishes probabilistic backlog, delay, and output bounds.

*Proof:* Let $A(0, t)$ and $D(0, t)$ be the arrival respectively departure process of the server. It follows from Def. 1 for any fixed sample path and all $t \geq 0$ that

$$\exists \tau \in [0, t] : D(0, t) \geq A(0, \tau) + S(\tau, t)$$
$$\Leftrightarrow \exists \tau \in [0, t] : A(0, t) - D(0, t) \leq A(0, t) - A(0, \tau) - S(\tau, t)$$

where $A(0, t) - D(0, t) = b(t)$ is the backlog at time $t$ and $A(0, t) - A(0, \tau) = A(\tau, t)$. Reformulation using min-plus de-convolution completes the proof of the backlog bound.

It follows from the definition of delay and Def. 1 for any fixed sample path and all $t \geq 0$ that

$$d(t) = \inf[s \geq 0 : A(0, t) - D(0, t + s) \leq 0]$$
$$\Rightarrow d(t) \leq \inf[s \geq 0 : A(0, t) - A(0, \tau) - S(\tau, t + s) \leq 0,$$
$$\forall \tau \in [0, t + s]].$$

Since $A(0, t) - A(0, \tau) \leq 0$ for $\tau \geq t$ and generally $S(\tau, t + s) \geq 0$ for all $\tau \in [0, t + s]$ it is sufficient to just consider all $\tau \in [0, t]$ where $A(0, t) - A(0, \tau) = A(\tau, t)$. Reformulation using min-plus de-convolution over the interval $[0, t]$ completes the proof of the delay bound.

Lastly, it follows from Def. 1 for any fixed sample path and all $t \geq s \geq 0$ that

$$\exists \tau \in [0, s]: D(0, s) \geq A(0, \tau) + S(\tau, s)$$
$$\Leftrightarrow \exists \tau \in [0, s]: D(0, t) - D(0, s) \leq D(0, t) - A(0, \tau) - S(\tau, s)$$
$$\Rightarrow \exists \tau \in [0, s]: D(0, t) - D(0, s) \leq A(0, t) - A(0, \tau) - S(\tau, s)$$

since $A(0, t) \geq D(0, t)$ for causality. With $D(0, t) - D(0, s) = D(s, t)$, $A(0, t) - A(0, \tau) = A(\tau, t)$ and using min-plus de-convolution the proof of the output bound is completed.

The respective probabilistic upper bound on the result of min-plus de-convolution of two statistically independent random processes $(A \oslash S)(s, t)$ follows from Lemma 2. ∎

After showing the basic network calculus theorems for dynamic servers the remainder of this section addresses schedulers which implement respective properties. Def. 1 is trivially fulfilled by deterministic servers [1]. Beyond this obvious class an example has been given in [36] for work-conserving servers with time-varying capacity and in [1], [36], [37] for the dynamic service curve based earliest deadline scheduler. The corresponding MGF of the offered service is, however, not derived in [1] and probabilistic performance bounds obtained from MGFs are provided in [23], [1] for the class of work-conserving constant rate servers only.

In the following we derive the leftover service under a general scheduling model which does not make assumptions about the order in which flows are served with respect to each other. Thus, it is conservative for most schedulers and it includes widely used aggregate first-in first-out scheduling as well as priority scheduling. The approach that is applied by deterministic network calculus is based on per-flow service curves which are derived by subtracting service that is consumed by cross-traffic [1], [38], [2]. When applying effective envelopes to model cross-traffic, probabilistic per-flow service curves arise naturally from deterministic servers [33], [32]. Similar results are obtained here for dynamic servers, however, using MGFs. The following derivations are made for two flows or aggregates of flows where lower indices denote flows respectively the service which is provided to a particular flow. Note that any of the two flows can actually consist of an arbitrary number of micro-flows.

**Proposition 1 (General Scheduling Model)** *Consider two flows with arrival processes $A_1(s, t)$ and $A_2(s, t)$ that are scheduled at a work-conserving server with service process $S(s, t)$ as defined in Ex. 1. Assume that $S(s, t)$ is nonnegative, increasing in $t$, and $S(s, s) = 0$. Flow two sees a dynamic server*

$$S_2(s, t) = \max[0, S(s, t) - A_1(s, t)].$$

*Assume $A_1(s, t)$ and $S(s, t)$ are statistically independent and have MGF $\mathsf{M}_{A_1}(\theta, s, t)$ and $\overline{\mathsf{M}}_S(\theta, s, t)$ respectively. The MGF of the dynamic server seen by flow two is upper bounded for $\theta \geq 0$ and $t \geq s \geq 0$ by*

$$\overline{\mathsf{M}}_{S_2}(\theta, s, t) \leq \min[1, \overline{\mathsf{M}}_S(\theta, s, t)\mathsf{M}_{A_1}(\theta, s, t)].$$

*Proof:* Let $A_i(0, t)$ and $D_i(0, t)$ be the arrival respectively departure process of flow $i \in [1, 2]$. For any fixed sample path and any $t \geq 0$ there exists $s = \sup[\tau \in [0, t] : D_i(0, \tau) = A_i(0, \tau), \forall i \in [1, 2]]$, that is $s$ is the beginning of the last backlogged period before $t$ if any or $s = t$ otherwise. For a work-conserving server and the defined $s$ it holds for any fixed sample path that

$$D_1(0, t) + D_2(0, t) = D_1(0, s) + D_2(0, s) + S(s, t)$$
$$= A_1(0, s) + A_2(0, s) + S(s, t).$$

With $D_1(0, t) \leq A_1(0, t)$ for causality we obtain

$$D_2(0, t) \geq A_2(0, s) + S(s, t) - (A_1(0, t) - A_1(0, s)).$$

Since $D_2(0, t) \geq D_2(0, s) = A_2(0, s)$ it follows that

$$D_2(0, t) \geq A_2(0, s) + \max[0, S(s, t) - (A_1(0, t) - A_1(0, s))].$$

With $A_1(0, t) - A_1(0, s) = A_1(s, t)$ the first part of the proof is completed. The MGF for statistically independent random processes $A_1(s, t)$ and $S(s, t)$ follows from (2) and (3). ∎

## IV. Scaling of End-to-End Performance Bounds

The scaling of end-to-end backlog and delay bounds in the number of servers in series, denoted by $n$, is a crucial aspect in network performance analysis. The benchmark set by deterministic network calculus for leaky bucket sources is a scaling in $\mathcal{O}(n^2)$ for summation of per-server bounds and $\mathcal{O}(n)$ for bounds obtained from end-to-end network service curves [26], [2]. This result highlights the importance of the concatenation theorem. Recently, probabilistic performance bounds that scale in $\mathcal{O}(n \log n)$ were obtained in [26] using effective network service curves and the exponentially bounded burstiness traffic model from [19]. We show that actually a scaling in $\mathcal{O}(n)$ can be achieved under the assumption of independence. Like [26] we consider the scenario shown in Fig. 1 where performance bounds are derived for the through flows that traverse all $n$ servers in series. Cross-traffic joins and leaves at each server where we make the assumption of statistical independence.

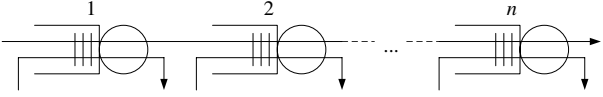

Fig. 1. Tandem servers with cross-traffic

We apply the $(\sigma(\theta), \rho(\theta))$-traffic model from [1] to derive closed-form solutions for end-to-end performance bounds. An arrival process $A(s,t)$ with MGF $\mathsf{M}_A(\theta,s,t)$ is $(\sigma(\theta), \rho(\theta))$-upper constrained for some $\theta > 0$ if for all $t \geq s \geq 0$ it holds that
$$\frac{1}{\theta} \ln \mathsf{M}_A(\theta, s, t) \leq \sigma(\theta) + \rho(\theta)(t-s). \tag{5}$$
A related quantity $1/(\theta t) \ln \mathsf{M}_A(\theta, 0, t)$ is known as effective bandwidth, for example [1], [34]. A wide variety of traffic models are known from related theories, including periodic sources, fluid sources, on-off sources, and regulated sources, see for example [1], [34], [28], [23], [16], [33] and references therein, which can be described using the $(\sigma(\theta), \rho(\theta))$-traffic characterization in (5).

Note that the right hand side of (5) depends only on the length of the interval $(s,t]$, that is $\delta = t-s$, but not on $s$ or $t$ itself. For univariate functions $x$ and $y$ with $x(t-s) = x(s,t)$ and $y(t-s) = y(s,t)$ for all $t \geq s \geq 0$ the operators $*$ and $\circ$ follow from (4) with $\delta = t-s \geq 0$ as
$$(x*y)(\delta) = \sum_{\tau=0}^{\delta} x(\delta - \tau) y(\tau)$$
$$(x \circ y)(\delta) = \sum_{\tau=0}^{\infty} x(\delta + \tau) y(\tau)$$
where we let $s \to \infty$ in case of the univariate $\circ$ operator. Note that the univariate convolution operator $*$ as opposed to its bivariate equivalent is commutative. We will use the $\circ$ operator also for $\delta < 0$ whereupon the sum is evaluated for all $\tau \in [-\delta, \infty)$.

The proof of Th. 3, which is shown in the sequel, uses the following lemma for composition of univariate $*$ and $\circ$ operators.

**Lemma 3 (Composition of univariate $*$ and $\circ$)** *Let $x(\delta)$, $y(\delta)$, and $z(\delta)$ be univariate, real-valued functions. It holds that*
$$(x \circ (y * z))(\delta) = ((x \circ y) \circ z)(\delta).$$

*Proof:* From the definition of univariate operators $*$ and $\circ$ it follows that
$$\begin{aligned}(x \circ (y*z))(\delta) &= \sum_{\tau_1=0}^{\infty} x(\delta + \tau_1) \sum_{\tau_2=0}^{\tau_1} y(\tau_1 - \tau_2) z(\tau_2) \\ &= \sum_{\tau_1=0}^{\infty} x(\delta + \tau_1) \sum_{u_1, u_2 \geq 0 : u_1 + u_2 = \tau_1} y(u_1) z(u_2) \\ &= \sum_{u_2=0}^{\infty} \sum_{u_1=0}^{\infty} x(\delta + u_1 + u_2) y(u_1) z(u_2) \\ &= ((x \circ y) \circ z)(\delta)\end{aligned}$$
which proves the composition rule. ∎

The following theorem extends the results from [23], [1] to servers with cross-traffic in series and shows the linear growth of end-to-end performance bounds in the number of tandem servers $n$. For ease of presentation we investigate the case of homogeneous servers and flows. Solutions for the general, heterogeneous case follow in a similar way.

**Theorem 3 (End-to-End Backlog and Delay Bounds)**
*Consider $n$ work-conserving, constant rate servers in series, each with capacity $C$ and independent $(\sigma_c(\theta), \rho_c(\theta))$ cross-traffic under the general scheduling model. Consider an independent $(\sigma(\theta), \rho(\theta))$ aggregate of through flows that traverses all $n$ servers in series. Any $b$ and $d$ are upper backlog respectively delay bounds with violation probability $\varepsilon \in (0,1]$ if for any $\theta \in (0, \infty)$ it holds that $C > \rho(\theta) + \rho_c(\theta)$ for stability and*
$$b \geq \sigma(\theta) + \frac{n\rho(\theta)\sigma_c(\theta)}{C - \rho_c(\theta)} + \frac{n \ln \gamma - \ln \varepsilon}{\theta}$$
$$d \geq \frac{\sigma(\theta)}{\rho(\theta)} + \frac{n\sigma_c(\theta)}{C - \rho_c(\theta)} + \frac{n \ln \gamma - \ln \varepsilon}{\theta \rho(\theta)}$$
*where* $\gamma = \frac{1-e^{-\theta\rho(\theta)(T+1)}}{1-e^{-\theta\rho(\theta)}} + \frac{1}{1-e^{-\theta(C-\rho(\theta)-\rho_c(\theta))}}$ *and* $T = \frac{\sigma_c(\theta)}{C-\rho_c(\theta)}$.

*For $n=1$ an improved condition under which $d$ is a delay bound is*
$$d \geq \frac{\sigma(\theta) + \sigma_c(\theta)}{C - \rho_c(\theta)} + \frac{\ln \gamma' - \ln \varepsilon}{\theta(C - \rho_c(\theta))}$$
*where* $\gamma' = \frac{1}{1-e^{-\theta(C-\rho(\theta)-\rho_c(\theta))}}$. *For $n \geq 2$ the conditions are*
$$d \geq \frac{\sigma(\theta) + n\sigma_c(\theta)}{C - \rho_c(\theta)} + \frac{n \ln \zeta - \ln \varepsilon}{\theta(C - \rho_c(\theta))}$$
*and* $d \geq \frac{ne^{-\theta(C-\rho(\theta)-\rho_c(\theta))}}{1-e^{-\theta(C-\rho(\theta)-\rho_c(\theta))}}$ *where* $\zeta = \frac{(1+d/n)^{1+d/n}}{(d/n)^{d/n}}$.

Clearly the conditions for backlog and delay bounds depend linearly on the number of servers in series $n$. The first two terms of the backlog bound can be interpreted as the burstiness of the through flows $\sigma(\theta)$ and the data accumulated due to the rate of the through flows $\rho(\theta)$ during the latency induced by cross-traffic $\sigma_c(\theta)/(C - \rho_c(\theta))$ at each of the $n$ servers. Accordingly, the delay bound comprises the latency $\sigma(\theta)/(C - \rho_c(\theta))$ which is due to the burst size of the through flows and the cumulated latency of the $n$ servers induced by cross-traffic, that is $n$ times $\sigma_c(\theta)/(C - \rho_c(\theta))$.

*Proof:* The end-to-end concatenation of $n$ statistically independent dynamic servers is described by convolution of the respective MGFs, see Th. 1. Using the univariate $*$ and $\circ$ operators with $\delta = t - s$ and applying Lemma 3 the bounds from Th. 2 build on the following MGF

$$\mathsf{M}_{A \oslash S}(\theta, \delta) \leq (\mathsf{M}_A(\theta) \circ (\overline{\mathsf{M}}_{S^1}(\theta) * \cdots * \overline{\mathsf{M}}_{S^n}(\theta)))(\delta) \\ = (\mathsf{M}_A(\theta) \circ \overline{\mathsf{M}}_{S^1}(\theta) \circ \cdots \circ \overline{\mathsf{M}}_{S^n}(\theta))(\delta). \quad (6)$$

The MGF of $(\sigma(\theta), \rho(\theta))$-constrained arrivals is upper bounded by $\mathsf{M}_A(\theta, \delta) \leq e^{\theta(\sigma(\theta) + \rho(\theta)\delta)}$ for $\theta > 0$. With Prop. 1 and using the notation $(x)^+ = \max[0, x]$ the MGF of the leftover service of a work-conserving, constant rate server with capacity $C$ and $(\sigma_c(\theta), \rho_c(\theta))$ cross-traffic is bounded according to

$$\overline{\mathsf{M}}_{S^i}(\theta, \delta) \leq e^{-\theta((C - \rho_c(\theta))\delta - \sigma_c(\theta))^+} = e^{-\theta(C - \rho_c(\theta))(\delta - T)^+}$$

where $T = \frac{\sigma_c(\theta)}{C - \rho_c(\theta)}$ is a latency induced by the cross-traffic. With (6) it follows for $(\sigma(\theta), \rho(\theta))$ through traffic that

$$\mathsf{M}_{A \oslash S}(\theta, \delta) \leq \sum_{u_n=0}^{\infty} \cdots \sum_{u_2=0}^{\infty} \sum_{u_1=0}^{\infty} e^{\theta(\sigma(\theta) + \rho(\theta)(\delta + u_1 + u_2 + \cdots + u_n))} \\ e^{-\theta(C - \rho_c(\theta))(u_1 - T)^+} e^{-\theta(C - \rho_c(\theta))(u_2 - T)^+} \cdots e^{-\theta(C - \rho_c(\theta))(u_n - T)^+}$$

and after some reordering we obtain

$$\mathsf{M}_{A \oslash S}(\theta, \delta) \leq e^{\theta(\sigma(\theta) + \rho(\theta)(\delta + nT))} \\ \prod_{i=1}^{n} \sum_{u_i=0}^{\infty} e^{-\theta((C - \rho_c(\theta))(u_i - T)^+ - \rho(\theta)(u_i - T))}.$$

The sums $\sum_{u_i=0}^{\infty}$ are divided into $\sum_{u_i=0}^{\lceil T \rceil - 1}$ and $\sum_{u_i = \lceil T \rceil}^{\infty}$ and solved for $\theta > 0$ and $\rho(\theta) > 0$ using geometric series which yields

$$\sum_{u_i=0}^{\lceil T \rceil - 1} e^{-\theta((C - \rho_c(\theta))(u_i - T)^+ - \rho(\theta)(u_i - T))} \\ = e^{-\theta \rho(\theta) T} \sum_{u_i=0}^{\lceil T \rceil - 1} \left(e^{\theta \rho(\theta)}\right)^{u_i} = e^{-\theta \rho(\theta) T} \frac{e^{\theta \rho(\theta) \lceil T \rceil} - 1}{e^{\theta \rho(\theta)} - 1} \\ = e^{-\theta \rho(\theta)(T+1)} \frac{e^{\theta \rho(\theta) \lceil T \rceil} - 1}{1 - e^{-\theta \rho(\theta)}} \leq \frac{1 - e^{-\theta \rho(\theta)(T+1)}}{1 - e^{-\theta \rho(\theta)}}$$

and under the stability condition $C > \rho(\theta) + \rho_c(\theta)$

$$\sum_{u_i = \lceil T \rceil}^{\infty} e^{-\theta((C - \rho_c(\theta))(u_i - T)^+ - \rho(\theta)(u_i - T))} \\ = e^{\theta(C - \rho(\theta) - \rho_c(\theta))T} \sum_{u_i = \lceil T \rceil}^{\infty} \left(e^{-\theta(C - \rho(\theta) - \rho_c(\theta))}\right)^{u_i} \\ = \frac{e^{-\theta(C - \rho(\theta) - \rho_c(\theta))(\lceil T \rceil - T)}}{1 - e^{-\theta(C - \rho(\theta) - \rho_c(\theta))}} \leq \frac{1}{1 - e^{-\theta(C - \rho(\theta) - \rho_c(\theta))}}$$

such that

$$\mathsf{M}_{A \oslash S}(\theta, \delta) \leq e^{\theta(\sigma(\theta) + \rho(\theta)(\delta + nT))} \\ \left(\frac{1 - e^{-\theta \rho(\theta)(T+1)}}{1 - e^{-\theta \rho(\theta)}} + \frac{1}{1 - e^{-\theta(C - \rho(\theta) - \rho_c(\theta))}}\right)^n. \quad (7)$$

From Th. 2 it follows that $b$ and $d$ are a backlog respectively delay bound which are violated at most with probability $\varepsilon$ if for any $\theta \in (0, \infty)$ it holds that

$$b \geq \frac{1}{\theta}(\ln \mathsf{M}_{A \oslash S}(\theta, 0) - \ln \varepsilon) \\ \frac{1}{\theta}(\ln \mathsf{M}_{A \oslash S}(\theta, -d) - \ln \varepsilon) \leq 0. \quad (8)$$

Inserting (7) into (8) and reordering completes the first part of the proof.

Note that the derivation of the delay bound evaluates larger intervals than stated in Th. 2 to simplify the derivation of a conservative, closed-form bound. Taking the corresponding restriction of the interval into account we have for $n = 1$ that

$$\mathsf{M}_{A \oslash S}(\theta, -d) \leq \sum_{\tau=d}^{\infty} e^{\theta(\sigma(\theta) + \rho(\theta)(\tau - d))} e^{-\theta((C - \rho_c(\theta))\tau - \sigma_c(\theta))} \\ = e^{\theta(\sigma(\theta) + \sigma_c(\theta) - \rho(\theta)d)} \sum_{\tau=d}^{\infty} \left(e^{-\theta(C - \rho(\theta) - \rho_c(\theta))}\right)^{\tau} \\ = e^{\theta(\sigma(\theta) + \sigma_c(\theta) - \rho(\theta)d)} \frac{e^{-\theta(C - \rho(\theta) - \rho_c(\theta))d}}{1 - e^{-\theta(C - \rho(\theta) - \rho_c(\theta))}}$$

where $\theta > 0$ and $C > \rho(\theta) + \rho_c(\theta)$ for stability. The delay bound follows immediately from (8).

For $n \geq 2$ we obtain from the first expression in (6) using the commutativity of the univariate $*$ operator that

$$\mathsf{M}_{A \oslash S}(\theta, -d) \leq \sum_{\tau=d}^{\infty} e^{\theta(\sigma(\theta) + \rho(\theta)(\tau - d))} \\ \sum_{u_{n-1}=0}^{\tau} \cdots \sum_{u_2=0}^{u_3} \sum_{u_1=0}^{u_2} e^{-\theta((C - \rho_c(\theta))u_1 - \sigma_c(\theta))} \\ e^{-\theta((C - \rho_c(\theta))(u_2 - u_1) - \sigma_c(\theta))} \cdots e^{-\theta((C - \rho_c(\theta))(\tau - u_{n-1}) - \sigma_c(\theta))}.$$

After some reordering we find

$$\mathsf{M}_{A \oslash S}(\theta, -d) \leq e^{\theta(\sigma(\theta) - \rho(\theta)d + n\sigma_c(\theta))} \\ \sum_{\tau=d}^{\infty} \sum_{u_i \geq 0 : \sum_{i=1}^{n} u_i = \tau} e^{-\theta(C - \rho(\theta) - \rho_c(\theta))\tau}.$$

With $q = e^{-\theta(C-\rho(\theta)-\rho_c(\theta))}$ the double sum becomes [39]

$$\sum_{\tau=d}^{\infty} \sum_{u_i \geq 0: \sum_{i=1}^{n} u_i = \tau} q^{\tau} = \sum_{\tau=d}^{\infty} \binom{\tau+n-1}{n-1} q^{\tau} \qquad (9)$$
$$= \frac{1}{p^n} \sum_{\tau=d}^{\infty} \binom{\tau+n-1}{n-1} p^n q^{\tau}.$$

Under the stability condition $C > \rho(\theta) + \rho_c(\theta)$ and for $\theta > 0$ we choose $p = 1 - q$ where $p, q \in (0,1)$ and give (9) a probabilistic interpretation: Consider a number of independent trials each having probability $p$ of being a success. The probability to obtain the $n$-th success exactly in the $(\tau+n)$-th trial is given by the negative binomial distribution. It equals $\binom{\tau+n-1}{n-1} p^n q^{\tau}$. Consequently, (9) can be interpreted as $1/p^n$ times the probability that $d+n$ or more trials are needed for $n$ successes.

The number of trials required to obtain $n$ successes is, however, $\sum_{i=1}^{n} X_i$ where the $X_i$ are independent geometric random variables that each describe the number of trials required to obtain only the $i$-th success [39]. That is $X_1$ is the number of trials required to obtain the first success, $X_2$ is the additional number of trials required to obtain the second success and so on. The probability that $x-1$ trials fail before a success occurs at the $x$-th trial is $\mathsf{P}\{X_i = x\} = pq^{x-1}$ for $x \geq 1$ and the corresponding MGF [40] becomes $\mathsf{M}_{X_i}(\vartheta) = \frac{pe^{\vartheta}}{1-qe^{\vartheta}}$ for $qe^{\vartheta} < 1$. The MGF of $X = \sum_{i=1}^{n} X_i$ follows as $\mathsf{M}_X(\vartheta) = (\mathsf{M}_{X_i}(\vartheta))^n$ and with Chernoff's bound we estimate

$$\frac{1}{p^n} \mathsf{P}\{X \geq d+n\} \leq \frac{1}{p^n} \inf_{\vartheta \in [0,\infty)} \left[ e^{-\vartheta(d+n)} \left( \frac{pe^{\vartheta}}{1-qe^{\vartheta}} \right)^n \right]$$

which is minimized at $e^{\vartheta} = \frac{d/n}{q(1+d/n)}$ if $d \geq \frac{nq}{1-q}$ to ensure $\vartheta \geq 0$. It follows that

$$\frac{1}{p^n} \mathsf{P}\{X \geq d+n\} \leq \left( \frac{q(1+d/n)}{d/n} \right)^d (1+d/n)^n$$
$$= \left( \frac{(1+d/n)^{1+d/n}}{(d/n)^{d/n}} \right)^n q^d$$

which provides an upper bound for (9). Assembling all parts we obtain

$$\mathsf{M}_{A \oslash S}(\theta, -d) \leq e^{\theta(\sigma(\theta)+n\sigma_c(\theta)-(C-\rho_c(\theta))d)}$$
$$\left( \frac{(1+d/n)^{1+d/n}}{(d/n)^{d/n}} \right)^n$$

and the delay bound follows by insertion into (8). ∎

## V. NUMERICAL RESULTS

We provide numerical delay bounds for $n$ servers with cross-traffic in series as shown in Fig. 1. The servers each have capacity $C$ and the leftover service is determined under the general scheduling model. We use stationary, single leaky bucket traffic sources with maximum burst size $b$ and rate $r$. For the MGF of such regulated sources, see for example [28], [16], it is known for $\theta \geq 0$ and $t \geq s \geq 0$ that $\mathsf{M}_A(\theta, s, t) \leq 1 + \frac{r(t-s)}{b+r(t-s)} \left( e^{\theta(b+r(t-s))} - 1 \right)$. We compute stationary delay bounds with violation probability $\varepsilon = 10^{-6}$ from Th. 1 and Th. 2 and compare the results to respective bounds obtained from deterministic network calculus, that is for $\varepsilon = 0$. Note that the probabilistic calculus recovers corresponding deterministic bounds for $\theta \to \infty$, whereas the optimal choice of $\theta$ effectively exploits the statistical multiplexing gain. We show a significant statistical gain for aggregation of independent flows, for independent cross-traffic, and for concatenation of independent servers, where the linear scaling of performance bounds in the number of traversed servers $n$ is confirmed.

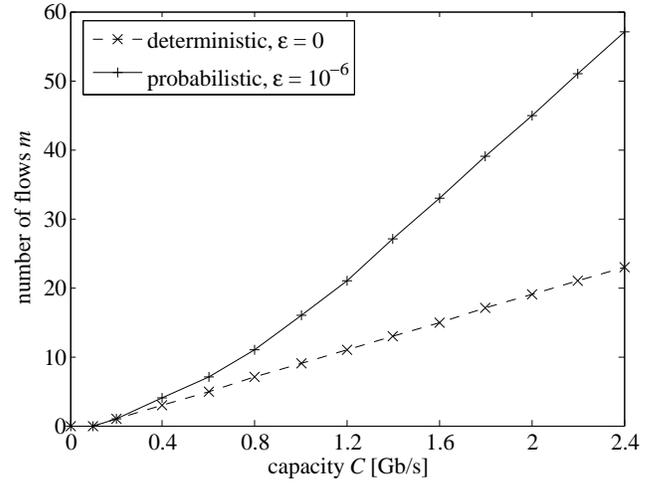

Fig. 2. Utilization under aggregate scheduling

Fig. 2 shows results for a single server which, however, may be a lumped, equivalent system that represents an entire network. For ease of presentation homogeneous flows with maximum burst size $b = 1$ Mb and rate $r = 30$ Mb/s are used. Flows are added unless the target delay bound $d \leq 10$ ms is violated. Clearly with increasing number of independent flows the probabilistic calculus can efficiently exploit statistical independence, such that the utilization can be significantly enhanced. For the investigated scenario the average utilization of a 2.4 Gb/s link can be increased from 0.29 to 0.71 by allowing for a small violation probability $\varepsilon = 10^{-6}$ of the target delay bound $d = 10$ ms.

Fig. 3 shows the impact of high priority cross-traffic at a priority scheduler. A constant rate server with capacity $C = 2.4$ Gb/s is analyzed and delay bounds are derived for an aggregate of 20 independent, low priority flows, each with leaky bucket parameters $b = 1$ Mb and $r = 30$ Mb/s. At the same time $m$ independent, high priority flows with leaky bucket parameters $b = 20/m$ Mb and $r = 600/m$ Mb/s are scheduled. Since the sum of the traffic parameters of all $m$ high priority flows is constant, the deterministic delay bound does not depend on $m$. It is 22.3 ms, whereas the probabilistic bound with violation probability $\varepsilon = 10^{-6}$ decreases from 15.3 ms for $m = 1$ and approaches 4.4 ms for large $m$.

Two effects which are due to statistical multiplexing can be itemized here: For $m = 1$ the difference between the deterministic and the probabilistic delay bound is almost entirely due to the statistical multiplexing gain among the 20 low priority flows. For an increasing number of high priority flows an additional, important statistical multiplexing gain among the $m$ high priority flows and between the two priority classes can be made accessible. For comparison the dotted lines at 2.9 ms and 4.4 ms show the probabilistic delay bound without cross-traffic and with constant bit rate cross-traffic respectively. Clearly as the number of independent high priority flows increases the aggregate of the high priority flows becomes smoother and for large $m$ the probabilistic delay bound approaches the bound obtained under constant bit rate cross-traffic.

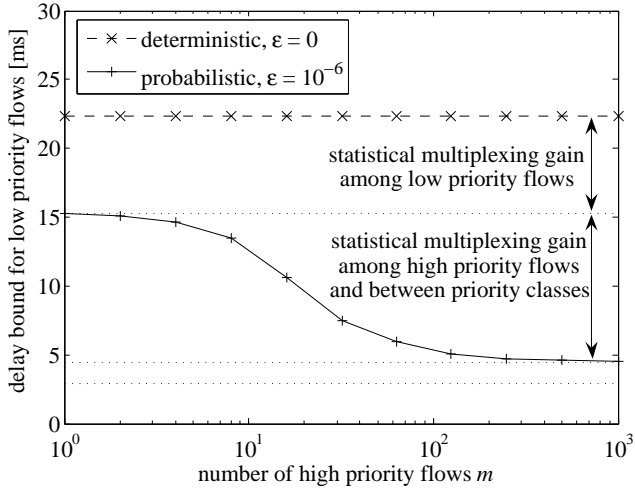

Fig. 3. Impact of high priority cross-traffic

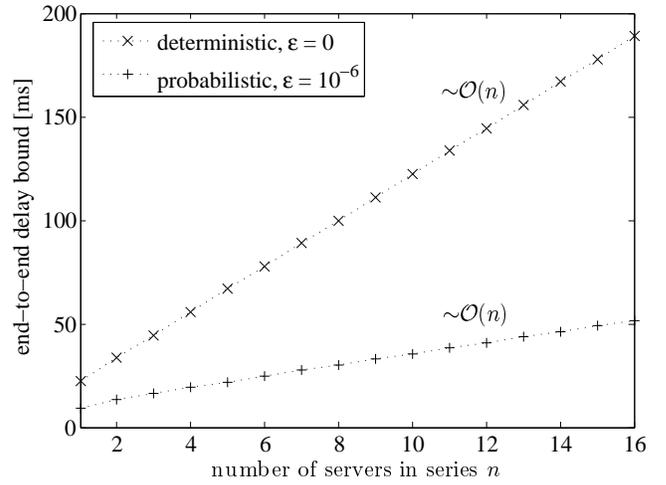

Fig. 5. End-to-end concatenation of servers in series

The statistical multiplexing gain becomes more and more apparent, if the number of flows is increased. To this end Fig. 4 shows results for the same scenario used to obtain Fig. 3, however, applying through and cross flows with burst size $b = 0.1$ Mb and rate $r = 3$ Mb/s. The number of through and cross flows are equal and are denoted by $m$. For the investigated 2.4 Gb/s link $m = 400$ corresponds to 400 through flows and 400 cross flows and results in a sustained network load of one. Fig 4 compares deterministic and probabilistic delay bounds for varying $m$ respectively varying network load. Clearly, if the utilization is very low there are few flows and only a small statistical multiplexing gain can be realized. Thus, the deterministic and probabilistic delay bounds are almost equal. On the other hand, if the utilization is very high, large delays are seen more frequently and the probabilistic delay bounds grow rapidly. Eventually, for a load of one the probabilistic calculus does not provide a finite delay bound, whereas the deterministic calculus still does. Between these extreme points the probabilistic calculus effectuates a significant statistical multiplexing gain resulting in a considerable reduction of delay bounds compared to the deterministic case.

Fig. 5 shows results for $n$ servers with cross-traffic in series. At each server the current cross-traffic is de-multiplexed and fresh, independent cross-traffic is multiplexed. The same parameters as used for Fig. 3 are applied and $m = 20$ is chosen. The results confirm the scalability of the approach. Both the deterministic and the probabilistic bounds scale in $\mathcal{O}(n)$, where $n$ is the number of servers in series. Thus, the important end-to-end scalability of deterministic network calculus has been achieved using a probabilistic calculus which effectuates a noticeable statistical multiplexing gain.

Further on, the so-called Pay Bursts Only Once phenomenon [2] known from deterministic network calculus can be observed in a probabilistic setting. In brief this phenomenon describes the observation that the increment of the end-to-end delay bound for each additional, concatenated server is smaller than the delay bound that would be obtained for the server in isolation.

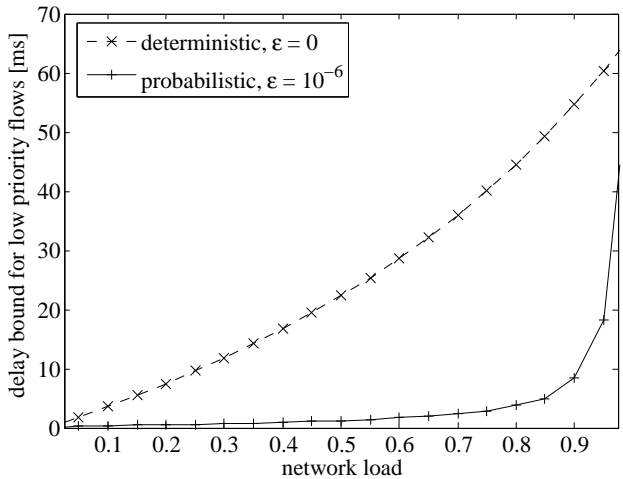

Fig. 4. Impact of the network load

## VI. Conclusions

We presented an end-to-end probabilistic network calculus with moment generating functions which efficiently utilizes statistical independence while preserving the intuitive convolution formulas and the fundamental concatenation theorem of deterministic network calculus. The derived framework meets the benchmark set by the deterministic calculus and achieves a linear scaling of end-to-end, per-flow backlog and delay bounds in the number of traversed servers.

We showed how moment generating functions can efficiently be used in the context of network calculus to exploit independence beyond statistical multiplexing of flows. A relevant gain is obtained for different scheduling disciplines, such as priority or first-in first-out scheduling, and for the concatenation of servers. We presented numerical results for tandem servers that support the efficiency of the approach.

## Acknowledgements

I gratefully acknowledge the valuable comments from A. Burchard, F. Ciucu, and J. Liebeherr who in particular called my attention to the scaling of end-to-end performance bounds. Also, I want to specially thank A. Burchard for pointing out the improved delay bound for $n \geq 2$ in Th. 3.